 \documentclass[article,5p,times,twocolumn]{elsarticle}

\usepackage{amssymb}

\graphicspath {{figures/}}

\usepackage{amsmath,amssymb,amsfonts}
\usepackage{algorithmic}
\usepackage{graphicx}
\usepackage{bm}
\usepackage{subfig}
\usepackage{caption}
\usepackage{balance}
\usepackage{mathtools}
\usepackage{booktabs}
\usepackage{placeins}
\usepackage{float}
\usepackage{changepage}
\usepackage{tabularx}
\usepackage{multirow}
\usepackage{makecell} 
\usepackage{bigstrut}
\usepackage{booktabs}
\usepackage{float}
\usepackage{orcidlink}
\usepackage{adjustbox}

\newcommand{\orcidNG}{0000-0002-8099-0627} 
\newcommand{\orcidJGM}{0000-0002-7422-5320} 
\newcommand{\orcidEV}{0000-0002-1627-6883} 
\newcommand{\orcidSLC}{0000-0003-3605-7351} 
\newcommand{\orcidMVO}{0000-0002-7680-3980} 
\newcommand{\orcidELP}{0000-0002-2743-1033} 
\newcommand{\orcidLCM}{0009-0009-3915-1707} 

\DeclareTextComposite{\.}{T1}{i}{`\i}
\DeclareTextComposite{\.}{T1}{\i}{`\i}
\newcommand{\RomanI}{I}

\journal{?}

\begin{document}

\begin{frontmatter}



\title{A  Novel Data Augmentation Tool for Enhancing Machine Learning Classification: A New Application of the Higher Order Dynamic Mode Decomposition for Improved Cardiac Disease Identification}


\author[inst1,inst2]{Nourelhouda Groun\orcidlink{\orcidNG}}
\ead{gr.nourelhouda@alumnos.upm.es}
\affiliation[inst1]{organization={ETSI Aeronáutica y del Espacio - Universidad Politécnica de Madrid},
            addressline={ Pl. del Cardenal Cisneros, 3}, 
            postcode={28040}, 
            state={Madrid},
            country={Spain}}

\affiliation[inst2]{organization={ETSI Telecomunicación - Universidad Politécnica de Madrid},
            addressline={Av. Complutense, 30}, 
            postcode={28040}, 
            state={Madrid},
            country={Spain}}

\author[inst3,inst4]{Mar\'{i}a Villalba-Orero\orcidlink{\orcidMVO}}
\ead{mvorero@ucm.es}
\affiliation[inst3]{organization={Departamento de Medicina y Cirugía Animal, Facultad de Veterinaria - Universidad Complutense de Madrid},
            addressline={Av. Puerta de Hierro}, 
            postcode={28040}, 
            state={Madrid},
            country={Spain}}
            
\affiliation[inst4]{organization={Centro Nacional de Investigaciones Cardiovasculares (CNIC) },
            addressline={C. de Melchor Fernández Almagro, 3}, 
            postcode={28029}, 
            state={Madrid},
            country={Spain}}

\author[inst3]{Luc\'{i}a Casado-Mart\'{i}n\orcidlink{\orcidLCM}}
\ead{lucasa04@ucm.es}
\author[inst4]{Enrique Lara-Pezzi \orcidlink{\orcidELP}}
\ead{elara@cnic.es}
\author[inst1,inst5]{Eusebio Valero \orcidlink{\orcidEV}}
\ead{eusebio.valero@upm.es}
\author[inst1,inst5]{Jes\'us Garicano-Mena \orcidlink{\orcidJGM}}
\ead{jesus.garicano.mena@upm.es}
\author[inst1,inst5]{Soledad Le Clainche \orcidlink{\orcidSLC}}
\ead{soledad.leclainche@upm.es}
\affiliation[inst5]{organization={Center for Computational Simulation (CCS)},
            addressline={Boadilla del Monte}, 
            postcode={28660}, 
            state={Madrid},
            country={Spain}}

\begin{abstract}
In this work, a data-driven, modal decomposition method, the higher order dynamic mode decomposition (HODMD),  is combined with a convolutional neural network (CNN) in order to improve the classification accuracy of several cardiac diseases using echocardiography images. The HODMD algorithm is used first as feature extraction technique for the echocardiography datasets, taken from both healthy mice and mice afflicted by different cardiac diseases (Diabetic Cardiomyopathy, Obesity, TAC Hypertrophy and Myocardial Infarction). A total number of 130 echocardiography  datasets are used in this work. The dominant features related to each cardiac disease were identified and represented by the HODMD algorithm as a set of DMD modes, which then are used as the input to the CNN. In a way, the database dimension was augmented, hence HODMD has been used, for the first time to the authors knowledge, for data augmentation in the machine learning framework. Six sets of the original echocardiography databases were hold out to be  used as unseen data to test the performance of the CNN. In order to demonstrate the efficiency of  the HODMD technique, two testcases are studied: the CNN is first trained using the original echocardiography images only, and second training the CNN using a combination of the original images and the DMD modes. The classification performance of the designed trained CNN shows that combining the original images with the DMD modes improves the results in all the testcases, as it improves  the accuracy by up to  22\%. These results show the great potential of using the HODMD algorithm as a data augmentation technique.
\end{abstract}




\end{frontmatter}


\section{Introduction}

In the health care industry, heart diseases are one of the most common cause of death globally \cite{virani2021heart}. Therefore,
it is important to use different tools to observe
the functional status of the heart.  Researchers and medical practitioner rely fundamentally on cardiovascular images for the clinical diagnosis of heart disease. These images provide crucial information for further clinical procedures, which makes it very important to take advantage of these images to  extract as much  information as possible and provide an accurate diagnosis. There is no doubt that researchers in the medical field are doing an outstanding job in diagnosing different heart disease, but we can not ignore the fact that Artificial intelligence (AI) and machine learning  (ML) have influenced every field of cardiovascular imaging through several applications.
Processing cardiac images and assisting in the diagnosis of different cardiac diseases through deep learning has become a popular topic. Researchers have introduced numerous applications of deep learning for cardiac images, such as detecting the abnormalities of regional left ventricular contraction and wall motion (\cite{parisi1981quantitative}, \cite{kusunose2020deep},\cite{shalbaf2013automatic}), viewpoint classification (\cite{gao2017fused}, \cite{yang2020deep},\cite{madani2018fast}) as well as heart segmentation (\cite{jang2018automatic},\cite{leclerc2019deep}) and noise and artifact reduction (\cite{wolterink2017generative},\cite{kang2019cycle},\cite{green20183}).\\
Another common application of deep learning for medical imaging is classification for cardiac diseases prediction. Several classification techniques have been proposed to help health-care professionals in diagnosing heart disease. Imayanmosha \textit{et al.} \cite{wahlang2021deep} used two deep leaning methodologies (a Recurrent Neural Network (RNN) based methodology (Long Short Term Memory (LSTM)) and an Autoencoder based methodology (Variational AutoEncoder (VAE)) to perform two different kinds of classification: (i) binary classification (normal or abnormal), (ii) categorical classification (for six different classes of regurgitation) using 2D echo images, 3D Doppler images, and videographic images. Both the methodologies did not score more than 80\% accuracy for the binary classification, however, they correctly classified the six classes of regurgitation with 96\% accuracy. Similarly,  Madani \textit{et al.} \cite{madani2018deep} developed two classifiers, a supervised model and a semi-supervised generative adversarial network, for the classification of left ventricle hypertrophy (LVH) and 15-view echocardiography images. The supervised model achieved 91.2\% and 94.4\% accuracy for LVH and view classification respectively, meanwhile, the semi-supervised model scored greater than 92.3\% for the LVH and greater than 80\% for the view classification. Tsai \textit{et al.} \cite{tsai2004medical} introduced a genetic-algorithm based fuzzy-logic approach to classify of myocardial heart disease from ultrasonic images. Beside their proposed method, the researchers investigated three other classification models (propagation learning method (BP–NN method), neural network with GA learning method (GA–NN method), and fuzzy method (without GA operation)). These methods were used to classify 90 samples of echocardiographic images from 45 subjects, achieving an accuracy of 82.1\%, 88.7\%, 91.4\% and 95.9\%, for BP–NN, GA–NN, fuzzy and GA-fuzzy, respectively, where the results showed the superiority of the GA-based fuzzy method. Meanwhile, Germain \textit{et al.} \cite{germain2021classification} evaluated different available convolutional neural networks (CNN) in classifying 1200 cine magnetic resonance (MR) sequences into three classes: normal, hypertrophic cardiomyopathy and dilated cardiomyopathy. The tested models gave quite similar results, achieving an accuracy of 98\% for the three cardiomyopathies classification.  \\

Even though all the previous researchers have approached cardiac diseases classification in different ways, they all aim towards the same goal, which is improving the classification accuracy.  In this work, we focus on the same goal. However, we explore the classification of  cardiac diseases using a new and  different approach than the previously mentioned techniques. In particular, a convolutional neural network is combined with a data-driven method named the higher order dynamic mode decomposition (HODMD) (\cite{le2017higher}), which has been introduced to the medical field quite recently (\cite{groun2022higher,groun2022novel,groun2023higher}), in order to classify echocardiography datasets into five categories: healthy, diabetic cardiomyopathy, obesity, myocardial infarction and TAC hypertrophy. In more details, what makes our approach different from the classical  techniques, is the fact that the classification is not done directly using the original echocardiography images, but instead, we perform a pre-processing step, where the HODMD algorithm is used to identify and extract the dominant features related to the different cardiac conditions. These features will be presented in a set of DMD modes, and will be used to augment the original database. In more details, a first testcase (testcase 01) is taken, where the build CNN is trained using a database consists of $10000$ original echocardiography images ($2000$ image per class). Next, a second testcase (testcase 02) is investigated, where
the training will be done using the combination of both the original images and the DMD modes. The performance of the CNN is also evaluated on a set of completely new, unseen data. The results, which will be detailed in the coming sections, show a clear improvement in the accuracy when using the DMD modes to augment the original database. Furthermore, employing the HODMD for the analysis of these databases saves time and effort of data preparation, while preserving a better classification accuracy. \\ 

The remaining of the article is organized as follows: section \ref{Methd} will introduce the methods used. The materials and data pre-processing are explained in section \ref{material}. The different results obtained are presented in section \ref{Results} and section \ref{concl} will hold the conclusions. 

\section{Methods} \label{Methd}

\subsection{Higher order dynamic mode decomposition}\label{Methodology}
Higher order dynamic mode decomposition (HODMD) is an extension of a data analysis tool, well established in the fluid mechanics field, named dynamic mode decomposition (DMD)~\cite{schmid2010dynamic}. HODMD \cite{le2017higher} was introduced for the analysis of complex data modeling non-linear dynamical systems.\\
The HODMD is a data-driven method and it operates on data  organized in matrix form as:

\begin{equation}\label{Eq1}
\bm{ V}_1^K=[\bm{v}_1,\bm{v}_2,\dots ,\bm{v}_k,\dots ,\bm{v}_K], 
\end{equation}

where $\bm{v_k}$ is a snapshot (e.g. an echocardiography frame) collected at time $t_k$, with $ k = 1,\dots , K $. Hence $\bm{ V}_1^K \in \mathbb{R}^{J \times K}$ , with $ J = N_x \times N_y$, where $  N_x \hspace{0.2cm} \textrm{and} \hspace{0.2cm} N_y$ are the total number of pixels on the $X$ and $Y$ directions, respectively. \\

The data can also be arranged in a tensor as:
\begin{equation}\label{Eq2}
\resizebox{.9\hsize}{!}{$\bm{X}_{x_1,x_2,k}  \hspace{0.1cm} \textrm{for} \hspace{0.1cm}  x_1=1,\dots ,N_x  ;  x_2=1,\dots ,N_y \hspace{0.1cm} \textrm{and} \hspace{0.1cm}  k=1,\dots ,K,$} 
\end{equation}
where $ x_1 $ and $ x_2 $ represent the position of each pixel in the frame plane containing the image, for the horizontal and vertical components and $K$ is the number of snapshots.\\

Similarly to DMD, the HODMD algorithm decomposes the signal into a number of modes $\bm{u}_m$,   as follows:
\begin{equation} \label{Eq002}
\resizebox{.8\hsize}{!}{$ \bm{v}(\bm{x},t_k)\simeq \sum\limits_{m=1}^M a_m\bm{u}_me^{(\delta_m+i\omega_m)t_k} \hspace{0.1cm} \textrm{for} \hspace{0.1cm} k = 1, \dots , K ,$} 
\end{equation}
where  $a_m$ are the amplitudes, $t$ is the time, with the frequencies $\omega_m$ and the growth rates $\delta_m$.


The HODMD algorithm can be summarized in the two following steps:\\

\begin{enumerate}
\item Dimensionality reduction: this step is achieved using the singular value decomposition (SVD), which will represent the snapshot matrix defined in eq. (\ref{Eq1}) as the product of three matrix factors: 
\begin{equation} \label{Eq3}
\bm{ V}_1^K \simeq \bm{W}\bm{\Sigma} \bm{T}^T  = \sum_{j=1}^{r = min(J,K)} \sigma_{j} \bm{w}_{j} \bm{t}_{j}^{T}, 
\end{equation}

where $\bm{W}$  and $\bm{T}$ are real, orthonormal matrices. The columns of $\bm{W}$ (noted $\bm{w}_j$) are the left singular vectors of $\bm{V}_1^K$ (related to spatial properties), and the columns of $\bm{T}$ (noted $\bm{t}_j$) are the right singular vectors of $\bm{V}_1^K$ (related to temporal properties). The matrix  $\bm{\Sigma} $ is a matrix with real, non negative entries on the diagonal and zeros off the diagonal, the elements of $\bm{\Sigma}$ (noted $\sigma_{j}$) are the singular values corresponding to the left and right singular vectors of $\bm{V}_1^K$, with $\sigma_1 \geq \dots \geq \sigma_j \geq \dots \geq \sigma_r$.\\

Finally, the tolerance $\varepsilon_{SVD}$ is used to determine the number $N$ of SVD modes to retain as 

\begin{equation} \label{Eq003}
\frac{\sigma_{N+1}}{\sigma_1} \leq \varepsilon_{SVD} .
\end{equation} 
Thus, Eq. (\ref{Eq3}) can be rewritten as
\begin{equation} \label{Eq4}
\bm{\widehat{V}}^K_1 = \bm{\Sigma} \bm{T}^T, \quad \textrm{with} \quad \bm{ V}^K_1 \simeq   \bm{W} \bm{\widehat{V}}_1^K . 
\end{equation}
where $\bm{\widehat{V}}^K_1$ will be called the \textit{reduced snapshot matrix}

\item The second step is called the DMD-d algorithm, which combines the standard DMD with Takens delay embedding theorem \cite{takens2006detecting}.  In this step, a $d$ lagged Koopman assumption is applied to the reduced snapshot matrix defined in eq. (\ref{Eq4}):
\begin{equation} \label{Eq5}
 \bm{\widehat{V}}^K_{d+1}\simeq \bm{\widehat{R}}_1\bm{\widehat{V}}^{K-d}_1+\bm{\widehat{R}}_2\bm{\widehat{V}}^{K-d+1}_2 + \dots + \bm{\widehat{R}}_d\bm{\widehat{V}}^{K-1}_d ,  
\end{equation}
where $ \bm{ \widehat{R}}_k= \bm{W}^T \bm{R}_k \bm{W}  $ are \textit{the reduced Koopman operators}.\\

The $d$ \textit{reduced Koopman operators} $ \bm{\widehat{R}}_1, \dots , \bm{\widehat{R}}_d$ in eq. (\ref{Eq5}) describe the dynamics of the system. All these operators are joined into one matrix, which will be called the  \textit{modified Koopman matrix} $\Tilde{R}$.  In order to obtain the components of the expansion eq. (\ref{Eq002}), the eigen decomposition of the modified Koopman matrix $\Tilde{R}$ is carried out, where its eigenvectors are used to calculate the DMD modes $\bm{u}_m$ and its eigenvalues are used to obtain the frequencies $\omega_m$ and growth rates $\delta_m$. Meanwhile, the amplitudes $a_m$ are calculated by least squares fitting. \\
The number of  DMD  modes to retain $M$ is determined using the tolerance $\varepsilon_{DMD}$ as 
\begin{equation}\label{Eq006}
\frac{a_{M+1}}{a_1}  \leq \varepsilon_{DMD}  
\end{equation}    
The algorithm is explained in detail in Ref. (\cite{le2017higher}) and the Matlab codes and more applications can be found in (\cite{vega2020higher}).
\end{enumerate}

\subsection{CNN architecture and training}
\begin{figure*}[h!]
    \centering
    \includegraphics[width=16cm, height=6cm]{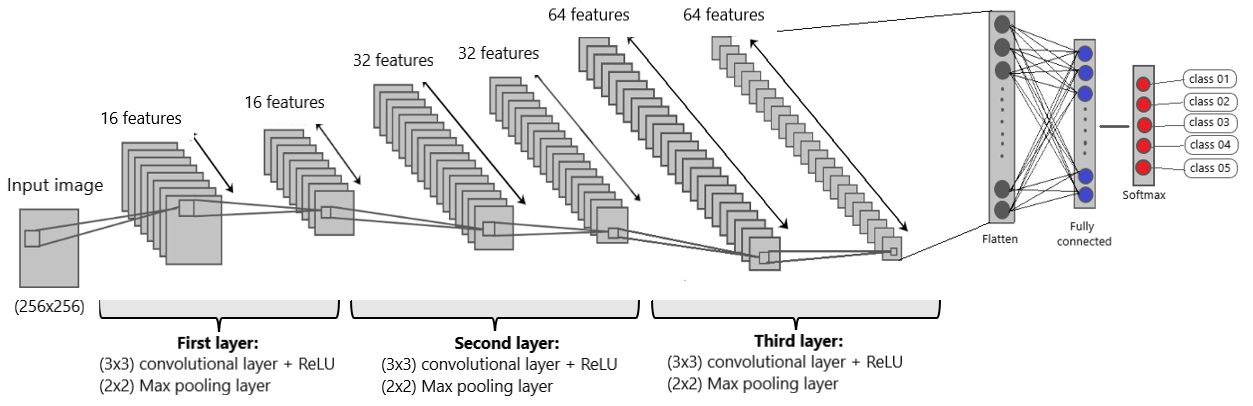}
    \caption{A sketch of the architecture of the Convolutional Neural Network which is designed to train a model for classifying the cardiac diseases.}
    \label{Fig01}
    
\end{figure*}

In this research, a convolutional neural network (CNN) have been designed for a categorical classification of echocardiography images. The CNN implementation was performed in Python, using the Keras library and TensorFlow backend. Hyperparameter tuning was performed to choose the ideal model architecture for our dataset, including the number of convolutional layers, the units of the dense layers and the learning rate. The training of the tuner went for 10 epochs with the max number of trials equals to 6. The model parameters, which were learned and used in building the CNN, are as follows: as seen in Fig. (\ref{Fig01}), the CNN is built using three ($3\times 3$) convolutional layers, where each one is followed by a ($ 2 \times 2$) max pooling layer. A flatten layer was included next and finally, two dense fully connected layer are added, such that, the first fully connected layer is accompanied with the activation function Rectified Linear Unit (ReLU) and the second one is accompanied with the soft-max activation function.

This architecture is considered a foundation for most of deep-learning-based classification models and it has proven to be one of the most efficient and unmatched in accuracy and robustness for image classification (\cite{lee2017deep,acharya2018automated,sharma2018analysis,cai2020review,puttagunta2021medical}). The parameters used for the training process are determined based on the size of the dataset. In our case, we first resize all the images to $256 \times 256$. Next, we train the model with a batch size of $128$,  optimizer RMSprop, learning rate (LR) of $10^{-3}$ (obtained using the hyper-parameter tuner), and categorical-crossentropy as loss function. Regarding the number of epochs,  "Early Stopping" using the “callbacks” argument is employed to prevent over-fitting. The monitor specified for this step is the validation accuracy, with patience argument equals to $40$: thus allows the training to continue for up to an additional $40$ epochs after the point that validation accuracy stops improving. Performance metrics are chosen to be validation loss and validation accuracy and finally,  the confusion matrix is used to determine the nature and the rate of misclassification.

\section{Material} \label{material}
\subsection{Echocardiography data}

This study included 130 echocardiography  datasets (videos) taken with respect to  a long axis view (LAX). A sample image is shown in Fig. \ref{Fig2} (Fig. (\ref{Fig2_a}): pre-cropping and Fig. (\ref{Fig2_b}): post-cropping). All the databases has a number of frames (which we call \textit{snapshots}) varying between 120-300. The 130 datasets split equally into 5 classes, each class is associated to  one of the following cases: healthy (H), obesity (Ob), diabetic cardiomyopathy (DC), myocardial infarction (MI) and TAC hypertrophy (HT). Each one of these classes encompass 26 sequences (samples): 20 of those are used for training; 100 successive frames (snapshots) of each sample are taken, resulting in 2000 images per class, which leads to a dataset with a total number of 10000 images (dataset 01). The final database is then divided into training, validation and testing/prediction (7000 images for training and 2500 for validation and 500 for prediction/testing). The remainder 6 samples (per class) are hold-out for testing as new, unseen data, where 90 successive images from each sample are taken, resulting in a second test database with a total number of 2700 images (testing \RomanI). This second testing database, which consists of original images only, is used for both testcases in order to conduct a fair comparison.

    

\subsection{Data pre-processing using the HODMD algorithm}
\begin{figure}[h!]

    \centering
\subfloat[ \label{Fig2_a}]{\includegraphics[width=3.5cm, height=3cm]{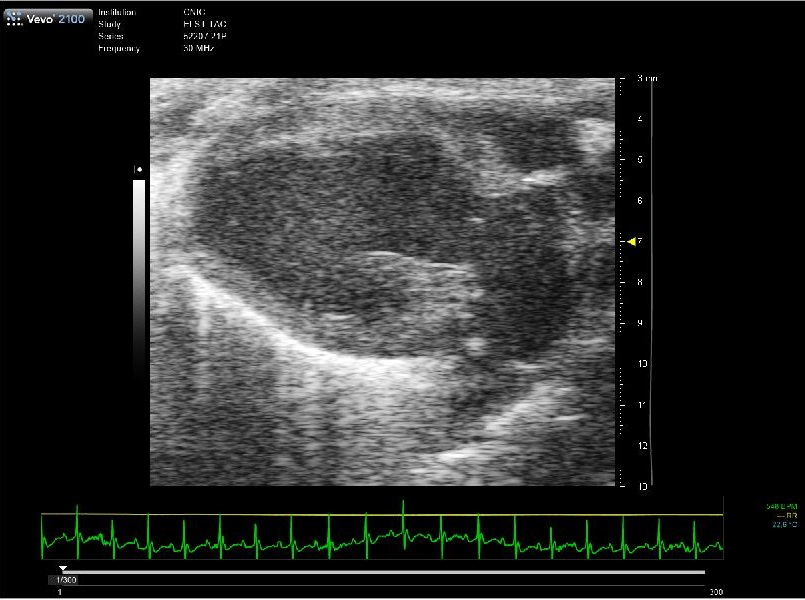}} \hspace{0.1cm} 
\subfloat[ \label{Fig2_b} ]{\includegraphics[width=3.5cm, height=3cm]{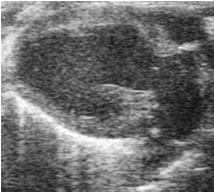}}  \\	
\caption{Original echocardiography image for LAX (a): before cropping, (b) after cropping}
    \label{Fig2} 

\end{figure}

In order to prepare the data for the classification process, a pre-processing stage was first carried out. This phase is accomplished in two steps:
\begin{itemize}
\item All the snapshots of each dataset were first imported and then cropped, where all the medical information displayed in the images are removed (as seen in Fig. (\ref{Fig2})). The cropped images of each dataset are then converted to gray scale images and arranged in an individual tensor (as in eq. (\ref{Eq2})). 
\item The prepared tensors are then analyzed separately using the HODMD algorithm implemented on Matlab\textsuperscript{\textregistered}~\cite{MATLAB:R2020b}. In this step the HODMD algorithm is used as a feature extraction technique, where for each dataset the HODMD works on identifying and extracting dominant features related to each one of the mentioned classes and represent them in sets of DMD modes (detailed explanation of this step can be found in our previous work \cite{groun2022higher}). All the obtained DMD modes are plotted and well observed in order to pick the modes, which will be employed for the augmentation process.
\item The criterion for the selection of the modes is related to the quality of the identified DMD modes. In particular, most representative modes (clear patterns), with high amplitudes and low noise levels, are chosen to augment the databases. Moreover, there are cases where the analysis of the echocardiography sequences results in either few DMD modes or modes with high noise levels. As a consequence, is these cases the average of the adequate DMD modes, which can be used for the augmentation, is 10 modes. Hence, To accomplish a balanced dataset for the classification, 10 DMD modes for each sample are taken, resulting in a total number of 200 DMD mode per class, this enables to enrich the original dataset 01 with 1000 additional images.
\item The additional 1000 images are split and arranged into training, validation and testing (prediction) folders, to be used to augment dataset 01, which leads to dataset 02. Hence, dataset 02 will have a total number of 7700 images for training, 2750 images for validation and 550 images for testing. 



\end{itemize}  

Table \ref{DATA} gathers the statistics for the datasets employed in this work.

\section{Results} \label{Results}
In the following, the results are elaborated using, first, the accuracy versus the number of epochs plots (e.g. Fig. \ref{Fig3_a}), and they display both the training and validation accuracies, which is considered the best metric, as model performance is best measured using accuracy as the metric of choice \cite{novakovic2017evaluation}. Second, the confusion matrix (e.g. Fig. \ref{Fig3_c}), which  evaluates the performance of the model when it make predictions either the testing or on the testing \RomanI~ (unseen dataset), as it shows the predictions of the model against the ground truth (true labels), giving the number of samples correctly classified in the diagonal and samples misclassified off the diagonal.

\renewcommand{\arraystretch}{1.5}
\begin{table*}
\begin{center}
\begin{tabular}{cc|c|c|c|c||c|||c|c|c||c|}
\cline{3-11}
 & & \multirow{2}{1.5cm}{Class}& \multicolumn{4}{c|||}{\centering Dataset 01: original images } 
& \multicolumn{4}{c|}{\centering Dataset 02: DMD-mode augmented }   
\bigstrut \\ \cline{4-11}
  & & &\multicolumn{1}{c|}{Training} & \multicolumn{1}{c|}{Validation} & \multicolumn{1}{c||}{Testing}  &\multicolumn{1}{c|||}{Testing \RomanI}  & \multicolumn{1}{c|}{Training} & \multicolumn{1}{c|}{Validation} & \multicolumn{1}{c||}{Testing} & \multicolumn{1}{c|}{Testing \RomanI} 


\bigstrut \\ \cline{2-11} 
\multicolumn{1}{ c|  }{\multirow{6}{*}{\rotatebox[origin=c]{90}{ 
 \hspace{-0.8cm} \footnotesize 4 classes classification   }} } & \multicolumn{1}{ c|  } {\multirow{5}{*}{ \rotatebox[origin=c]{90}{\hspace{-0.6cm}  \footnotesize 5 classes classification}}} & 
 H & 1400 & 500 & 100 & 540 & 1540 & 550 & 110 & 540   \bigstrut \\ 
\cline{1-1} \cline{3-11}
\multicolumn{1}{ |c| }{} &  & DC & 1400 & 500 & 100 & 540 &  1540 & 550 & 110 & 540   \bigstrut \\
\cline{3-11}
\multicolumn{1}{ |c| }{}  &   & MI & 1400 & 500 & 100 & 540 & 1540 & 550 & 110  & 540 \bigstrut \\
\cline{3-11}
\multicolumn{1}{ |c| }{}  &   & Ob & 1400 & 500 & 100 & 540 & 1540 & 550 & 110 & 540\bigstrut \\
\cline{3-11}
\multicolumn{1}{ |c| }{}  &   & HT & 1400 & 500 & 100 & 540 &  1540 & 550 & 110 & 540    \bigstrut \\
\cline{1-11}

\end{tabular}
\caption{Number of images used for each testcase for both 4 and 5 classes classification. The acronyms used are  H: healthy, DC: diabetic cardiomyopathy, MI: myocardial infarction, Ob: obesity, HT: TAC hypertension.} \label{DATA}
\end{center}
\end{table*}

\subsection{Five-class classification experiment:}

\begin{figure*}[ht!] 

    \centering
\subfloat[ \label{Fig3_a}]{\includegraphics[width=7cm, height=6cm]{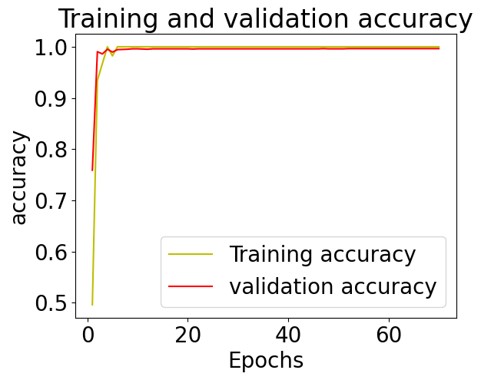}} \hspace{1cm}
\subfloat[ \label{Fig3_b}]{\includegraphics[width=7cm, height=6cm]{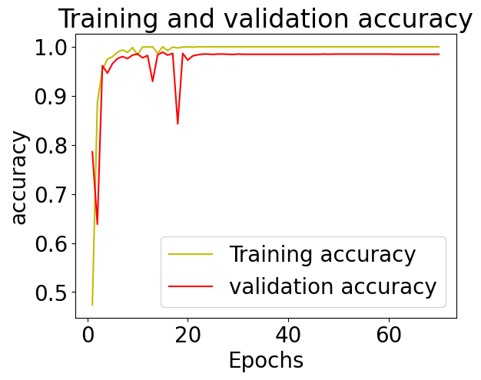}} \\

\subfloat[ \label{Fig3_c} ]{\includegraphics[width=8cm, height=7cm]{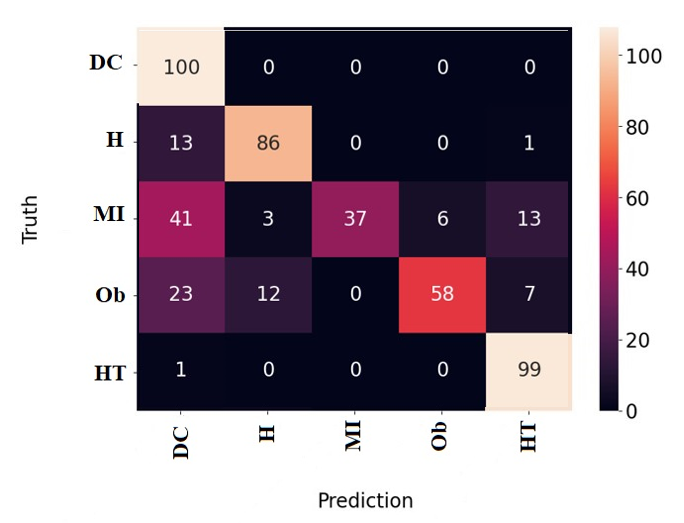}}  \quad 
\subfloat[ \label{Fig3_d} ]{\includegraphics[width=8cm, height=7cm]{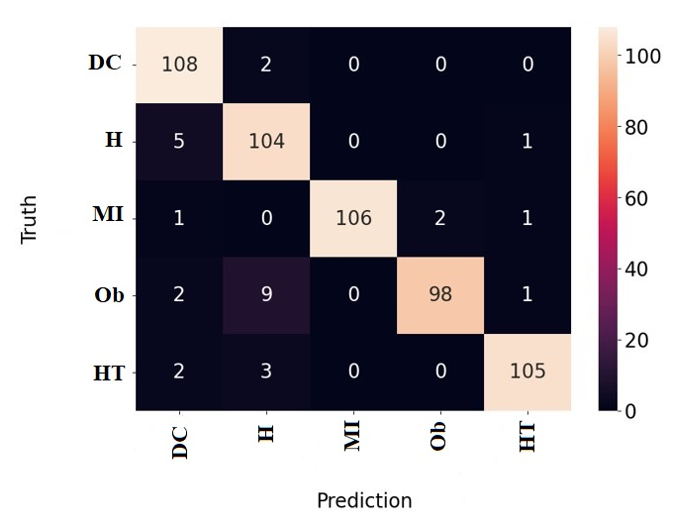}} \\

\subfloat[ \label{Fig3_e} ]
{\includegraphics[width=8cm, height=7cm]{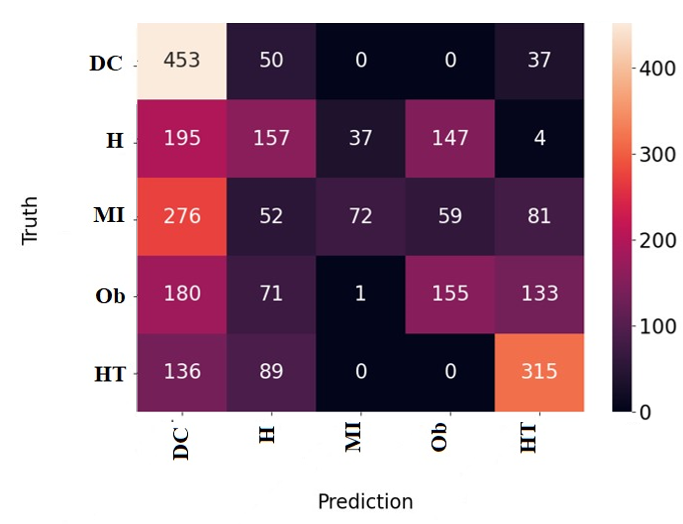}} \quad 
\subfloat[ \label{Fig3_f} ]{\includegraphics[width=8cm, height=7cm]{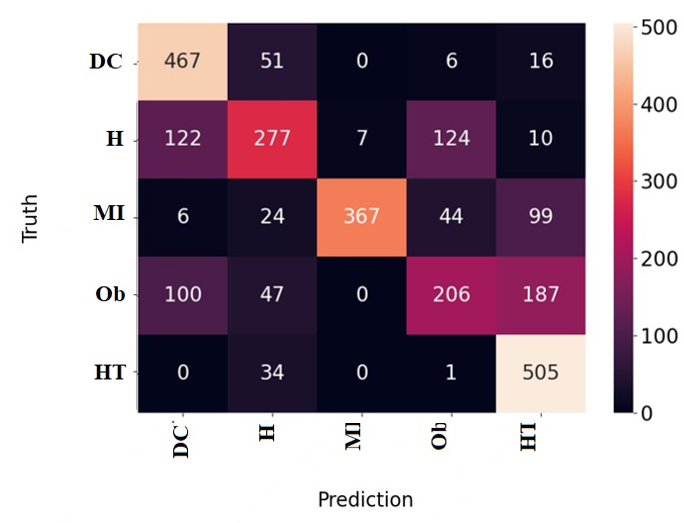}}  

\caption{Five-class classification experiment results. Training and validation accuracies throughout the training phase  and the confusion matrices displaying the perfprmance of the model on the testing and unseen data (testing \RomanI). Figs a, c and e refer to testcase 01, whereas Figs b, d and f refer to testcase 02. H: healthy, DC: diabetic cardiomyopathy, MI: myocardial infarction, Ob: obesity, HT: TAC hypertension.}
    \label{Fig3} 

\end{figure*}

The results obtained from the classification of all the five datasets re presented first.\\ 
As mentioned before, two different testcases has been carried out. In the first testcase, the CNN is trained using the original images only. Figure (\ref{Fig3_a}) shows the validation accuracy during the training process, as the validation accuracy  reaches and stabilizes at 99\% starting from approximately the 10th epoch.
However, as we can observe in Fig. (\ref{Fig3_c}) the CNN is having some difficulties achieving good results during the prediction phase, as the highest accuracy reached is 76\%, with 114 misclassifications out of 500.\\

In the second testcase, the database used for training is the DMD-augmented database (dataset 02). As seen in table (\ref{DATA}), each class is now represented by a  combination of $2000$ original images and $200$ DMD mode split between training, testing and validation. The results of this testcase are shown in Fig. (\ref{Fig3_b}) and (\ref{Fig3_d}). As we observe The validation accuracy is set to be 98\% starting from approximately the 25th epoch. Meanwhile, testing the performance on the testing  dataset increased the accuracy from  76\% to 94\%, with only 29 misclassifications out of 550. \\
The CNN is also tested on a set of new, unseen echocardiography images (testing \RomanI). As seen in Fig. (\ref{Fig3_e}), the CNN is still facing some difficulties achieving good results during the prediction when the training of the CNN is done using the original images only, as the highest accuracy reached is  42\%, with 1863 misclassifications out of 2700. Meanwhile, training the CNN using the augmented database increased the accuracy to 67\%, correctly classifications 1822 out of 2700 as seen in Fig. (\ref{Fig3_f}) (a summary of the results can be seen in Table \ref{LAX_Class_Sum}).  \\

Some of both original images and DMD modes have been included in the appendix in order to justify these results. As can be observed in Fig.(\ref{Fig06}), the original images are very similar, meaning that when splitting the database into training, testing and validation, the CNN will easily classify the validation set, resulting (as anticipated) in high validation accuracy. However, when the CNN is put through a more difficult task, which is classifying the unseen data in testing \RomanI, it fails to give satisfactory results. Meanwhile the DMD modes (seen in Fig. (\ref{Fig07})) can be seen displaying more details, enhancing  features and patterns, not easily perceptible in the original images. The clear difference in the DMD mode explains the slight drop in the validation accuracy when combining the original images with the DMD modes for the training, as the CNN is given a trickier assignment. Nevertheless, handing the CNN a more complicated mission definitely resulted a clear improvement in the accuracy when testing with the unseen data.


\subsection{Four-class classification experiment:}
\begin{figure*}[ht!] 

    \centering
\subfloat[ \label{Fig4_a}]{\includegraphics[width=7cm, height=6cm]{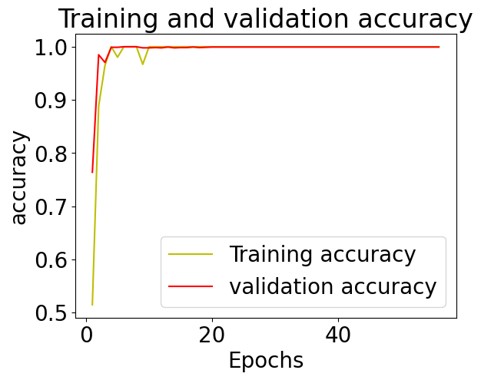}} \hspace{1cm}
\subfloat[ \label{Fig4_b}]{\includegraphics[width=7cm, height=6cm]{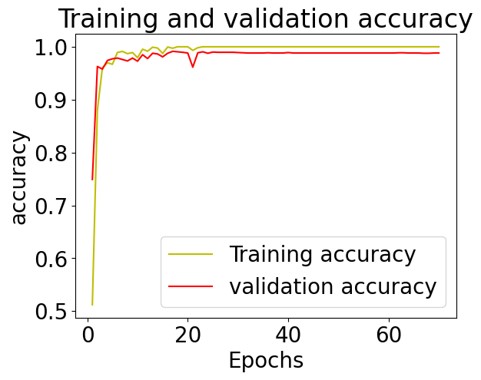}} \\

\subfloat[ \label{Fig4_c} ]{\includegraphics[width=8cm, height=7cm]{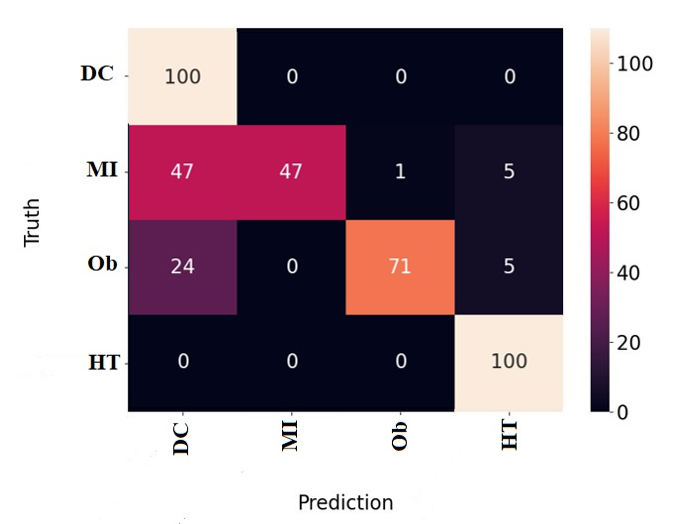}}  \quad 
\subfloat[ \label{Fig4_d} ]{\includegraphics[width=8cm, height=7cm]{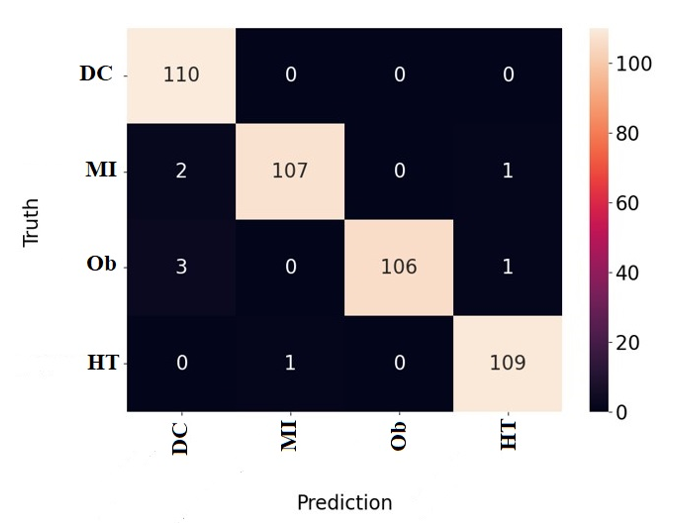}} \\

\subfloat[ \label{Fig4_e} ]
{\includegraphics[width=8cm, height=7cm]{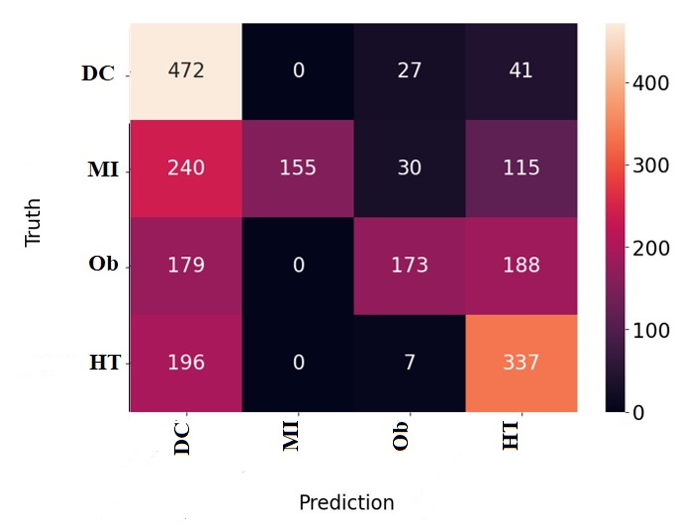}} \quad 
\subfloat[ \label{Fig4_f} ]{\includegraphics[width=8cm, height=7cm]{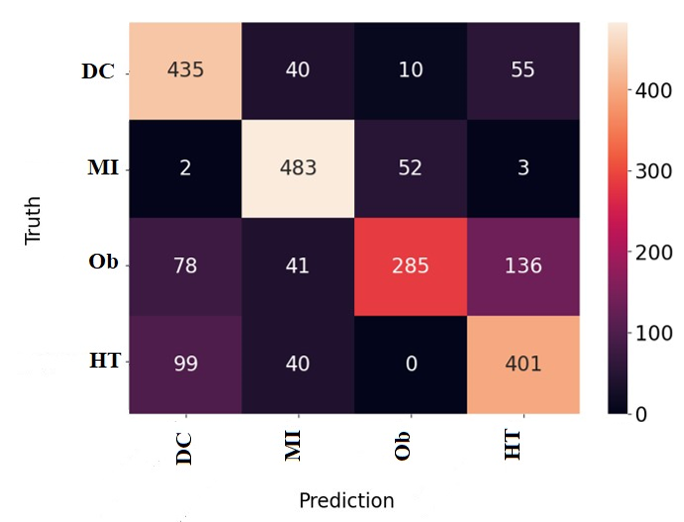}}  

\caption{Four-class classification experiment results. Training and validation accuracies throughout the training phase  and the confusion matrices displaying the perfprmance of the model on the testing and unseen data (testing \RomanI). Figs a, c and e refer to testcase 01, whereas Figs b, d and f refer to testcase 02. DC: diabetic cardiomyopathy, MI: myocardial infarction, Ob: obesity, HT: TAC hypertension.}
    \label{Fig4} 

\end{figure*}

It is well known from medical practice that it is not trivial even for medical doctors to discern healthy echocardiography data from unhealthy data. Moreover, one should bear in mind that 20 samples may not be an amount of data descriptive enough to properly train a CNN model. In an ideal world, we would resort to larger databases so that the CNN performance improves. However, well curated echocardiography data is difficult and expensive to generate. Therefore, in this section we propose an alternative experiment that will hopefully help us to assess the performance of the CNN classification of the HODMD augmented database. In particular, the healthy datasets are excluded in this experiment and the classification is focused on the four  cardiac diseases. \\
Similarly to the previous case, we carry out the same protocol. In the first testcase, a database consisting of $2000$ images per class is used for the training of the CNN. As seen in Fig. (\ref{Fig4_a}), the accuracy during the training process is set to be 99\%, which is anticipated. Meanwhile examining the performance of  the CNN on the testing dataset gave the accuracy of 79\%, with 82 misclassifications out of 400.\\
Next, for the second testcase, the combination of the original images and the DMD modes is utilized for the training of the CNN. As can be seen in Fig. (\ref{Fig4_b}), the accuracy during training reached 98\% starting from the 25th epoch and the prediction on the testing data increases the accuracy by 19\% reaching 98\%, misclassifying only 8 out of 440.\\
Equivalently to the five-class classification experiment, the CNN is tested on a set of unseen data in this case as well (testing \RomanI). Following the training of the CNN on both testcases,  the prediction on the unseen data (testing \RomanI) provided an accuracy of 52\%, with 1023 misclassifications out of 2160 for the first testcase and once again we notice a clear improvement in the accuracy with  74\%, misclassifying 556 out of 2160 in the second testcase, as can be seen in Figures  (\ref{Fig4_e}) and (\ref{Fig4_f}), respectively (see the summary of the results in Table \ref{LAX_Class_Sum}).\\

This second experiment confirms the initial hypothesis that the classification accuracy can be markedly improved by augmenting the training dataset with about 10\% of its size with HODMD modes, demonstrating the ability of using the HODMD algorithm as a data augmentation technique to upgrade the performance of machine learning approaches. 

\FloatBarrier


\begin{table}[!ht]
\begin{center}
\begin{adjustbox}{max width=0.5\textwidth}
\begin{tabular}{l l l l l}
\toprule
\multirow{2}{5cm}{Accuracy}& \multicolumn{2}{l}{\centering 05 Classes:}  & \multicolumn{2}{l}{\centering 04 Classes:} \\
\cline{2-5} & \multicolumn{1}{l}{Dataset 01} & \multicolumn{1}{l}{Dataset 02} & \multicolumn{1}{l}{Dataset 01} & \multicolumn{1}{l}{Dataset 02} \bigstrut \\ \hline
Validation  & 99\%   & 98\%  &     99\%  & 98\%   \bigstrut \\ \hline
Testing  & 76\% & 94\% & 79\% & 98\%   \bigstrut \\ \hline
Testing \RomanI~(Unseen data)  & 42\% & 67\% & 52\% & 74\% \bigstrut  \\ 
\bottomrule
\end{tabular} 
\end{adjustbox}
\caption{Accuracy results summary for all the experiments.}
\label{LAX_Class_Sum}
\end{center}
\end{table}

\section{Conclusions} \label{concl}
In this work, we have explored the use of a data-driven technique, the higher order dynamic mode decomposition (HODMD)  as a pre-processing tool and as a data augmentation technique, with the aim of improving the classification accuracy of five different cardiac conditions. In particular, the HODMD algorithm is combined with a convolutional neural network (CNN), to ameliorate the classification accuracy of five different classes into: healthy,  diabetic cardiomyopathy, obesity, TAC hypertrophy and myocardial infarction.  Ahead of the classification process, the HODMD is used to analyze all the echocardiography  datasets in order to identify and extract the features related to the different cardiac diseases. The extracted features are represented as what we call DMD modes.  The investigation conducted in this contribution involves the use of 130 echocardiography datasets, subdivided into 26 datasets (or samples) per class. In order to produce a balanced database for the classification, 20 samples from each class are taken for the training process, meanwhile the remaining 6 samples, are left-out to be used as unseen data for prediction. 
Two different testcases have been investigated in this contribution: (i) training the CNN with the original echocardiography images only, where 2000 images for each class (100 successive images per sample) are taken for the training process. The database for this testcase is divided into 7000 images for training, 2500 for validation and 500 for testing. (ii)  A database combining the original images and extracted features (DMD modes) is used for the training of the CNN. The total number of images used for this testcase is 7700 for training, 2750 for validation and 550 for testing. Meanwhile, for the prediction on new, unseen data, 90 original echocardiography images were taken from each one of the 6 hold-out datasets, resulting in a second test database with 2700 images, which will be used to evaluate the performance of the model in both testcases.  The results obtained from the different tescases exhibit the following: during the training of the CNN for both testcases, the achieved validation accuracy is comparable, as it reached 99\% in the first testcase and 98\% in the second one. However, when investigating the performance of the CNN on the prediction datasets, we notice a clear improvement in the accuracy when using the augmented database for the training, as the accuracy increased from 76\% to 94\% when examining using the testing dataset and from 42\% to 67\% when evaluating with the new, unseen data.\\
 
Another experiment was explored in this work, which is excluding the healthy datasets and focusing on the classification of the different pathologies. This experiment is considered based on several aspects, including medical experts opinions, the limited amount of data and the results we obtained from previous research. The additional experiment covers both the previous testcases as well. The results obtained display the same behavior of the five classes classification case, as we clearly see an improvement up to 22\% in the classification accuracy when employing the HODMD for augmenting the original database with the DMD modes. Furthermore, it is worth to mention that we have conducted a separate experiment where a higher number of DMD modes is employed for the augmentation. The results obtained from this experiment did not offer any significant improvement, which demonstrates that the correct criterion is indeed the quality of the DMD modes and not the number. Hence, choosing the correct criterion grants comparable results, permits the reduction of data and computations and saves time and effort.\\
Based on the results we observe in this contribution, we can conclude the efficiency of the HODMD algorithm as a data  augmentation technique. Combining the original images and the DMD modes enhances the performance of the CNN and its ability to predict on new unseen data, which is the main scoop of this research. Achieving higher accuracy levels is not the main aim of this investigation, however, working on improving these results is highly considered for future work.\\
\section{Acknowledgments:}
This publication has been supported by the grants  PID2021-124629OB-I00, TED2021-129774B-C22, TED2021-129774B-C21 and  PLEC2022-009235 funded by the Ministry of Science and Innovation (MCIN/AEI/ 10.13039/501100011033) and by the European Union “NextGenerationEU”/PRTR (“Plan de Recuperación, Transformación y Resiliencia de España”) and by FEDER to E.L-P, by grant PEJ-2019-TL/BMD-12831 from Comunidad de Madrid to E.L-P and by M.V.-O. by a Juan de la Cierva Incorporación Grant (IJCI-2016-27698) to M.V-O. “The CNIC is supported by the Instituto de Salud Carlos III (ISCIII), the Ministerio de Ciencia e Innovación (MCIN) and the Pro CNIC Foundation), and is a Severo Ochoa Center of Excellence (grant CEX2020-001041-S funded by MICIN/AEI/10.13039/501100011033)”. SLC acknowledges the grant PID2020-114173RB-I00 funded by MCIN/AEI/10.13039/501100011033 and the support of Comunidad de Madrid through the call Research Grants for Young Investigators from Universidad Polit\'ecnica de Madrid. Acknowledgments also to Agencia Española de investigation through  NextSim/AEI/10.13039/501100011033 and H2020, GA-956104.


\clearpage
\newpage
\appendix
\onecolumn


\begin{figure}
\textbf{Appendix A. Echocardiography frames and DMD modes visualization } \\
\vspace{0.5cm}

	\centering
	\textbf{\hspace{0.3cm} Dataset 01 \hspace{1cm}  Dataset 02 \hspace{1.2cm}  Dataset 03  \hspace{1.2cm}  Dataset 04 \hspace{1.2cm}  Dataset 05 }\\

\rotatebox{90}{\hspace{0.8cm}\textbf{H}}\hspace{0.0001cm}
\subfloat[  ]{\includegraphics[width=2.7cm, height=2.7cm]{LAX_H_Org}} \hspace{0.1cm} 
\subfloat[ ]{\includegraphics[width=2.7cm, height=2.7cm]{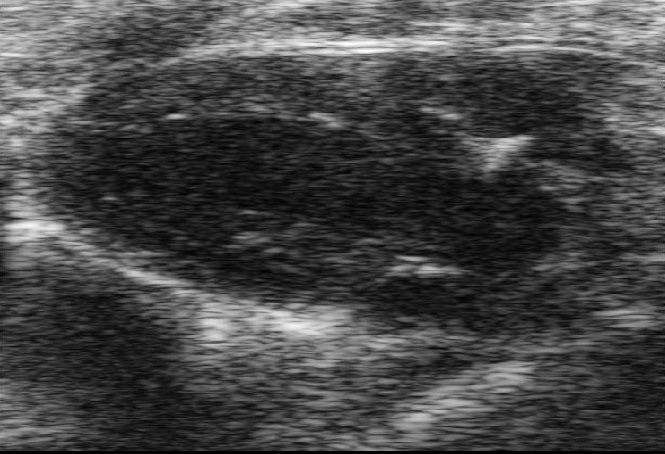}}\hspace{0.1cm} 
\subfloat[  ]{\includegraphics[width=2.7cm, height=2.7cm]{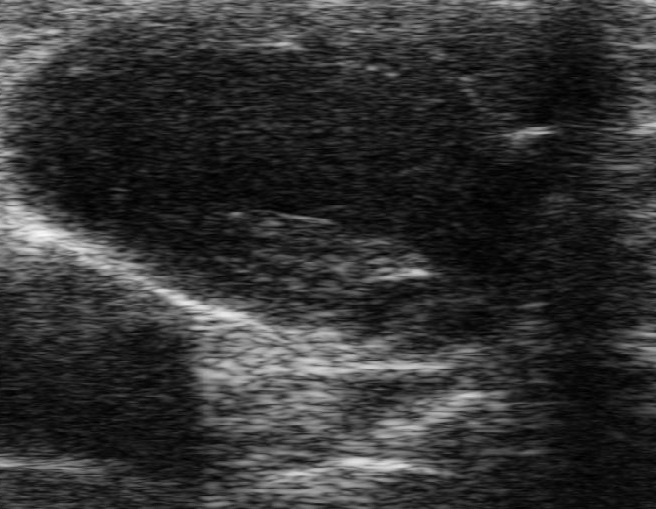}}\hspace{0.1cm}
\subfloat[ ]{\includegraphics[width=2.7cm, height=2.7cm]{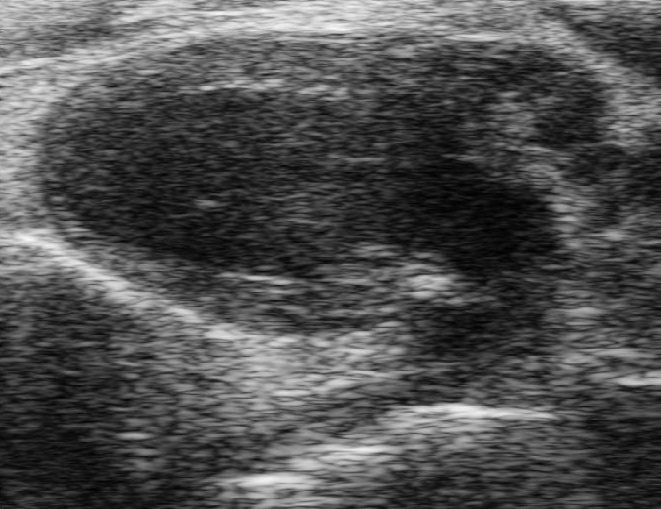}}\hspace{0.1cm} 
\subfloat[  ]{\includegraphics[width=2.7cm, height=2.7cm]{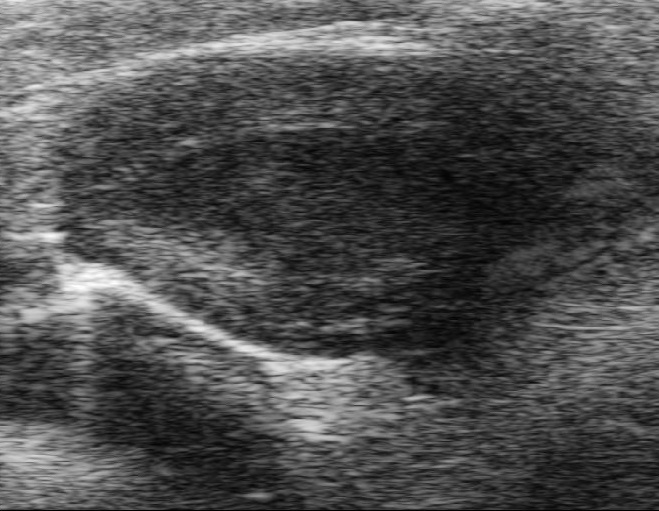}}\\

\rotatebox{90}{\hspace{1.1cm}\textbf{DC}}\hspace{0.1cm}
\subfloat[  ]{\includegraphics[width=2.7cm, height=2.7cm]{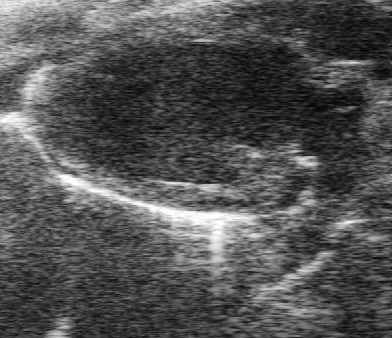}} \hspace{0.1cm} 
\subfloat[ ]{\includegraphics[width=2.7cm, height=2.7cm]{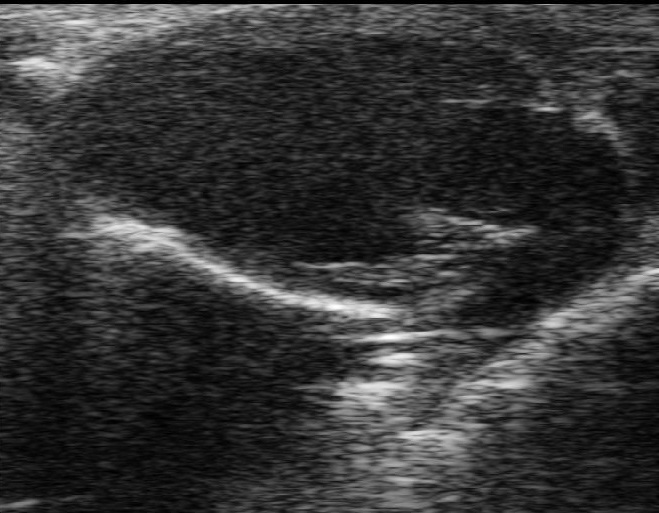}}\hspace{0.1cm}
\subfloat[ ]{\includegraphics[width=2.7cm, height=2.7cm]{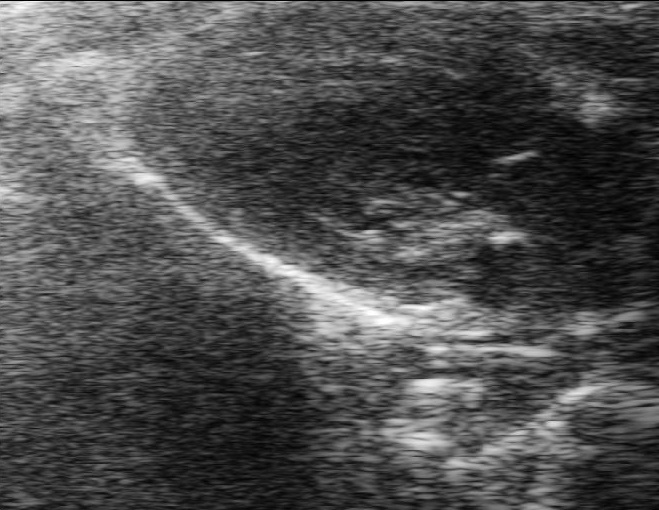}}\hspace{0.1cm}
\subfloat[ ]{\includegraphics[width=2.7cm, height=2.7cm]{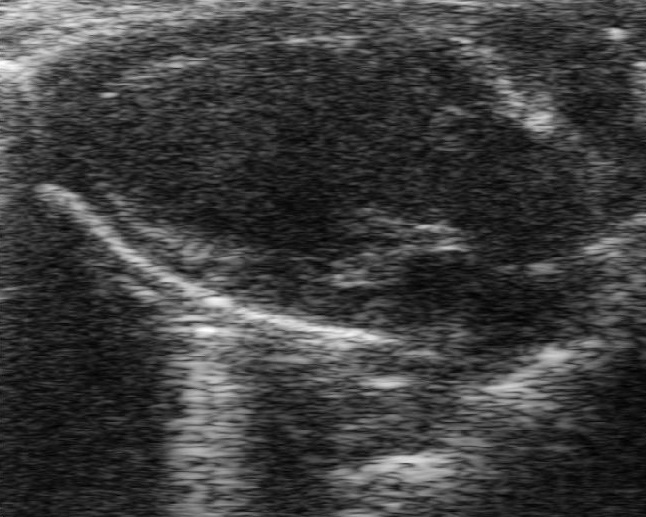}}\hspace{0.1cm}
\subfloat[ ]{\includegraphics[width=2.7cm, height=2.7cm]{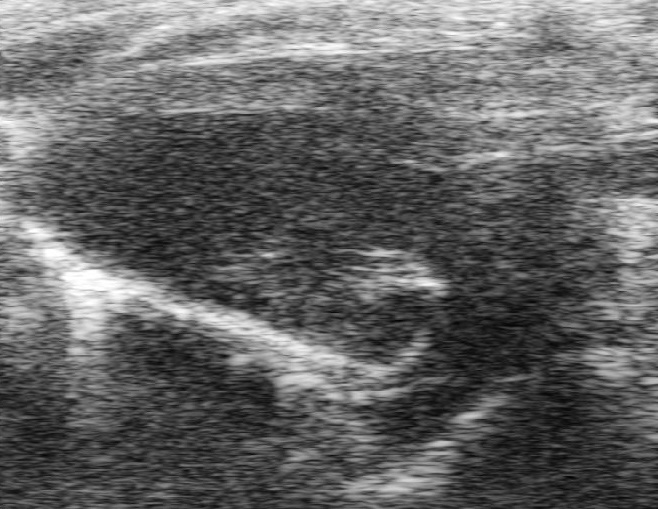}}\\

\rotatebox{90}{\hspace{1.1cm}\textbf{MI}}\hspace{0.1cm}
\subfloat[ ]
{\includegraphics[width=2.7cm, height=2.7cm]{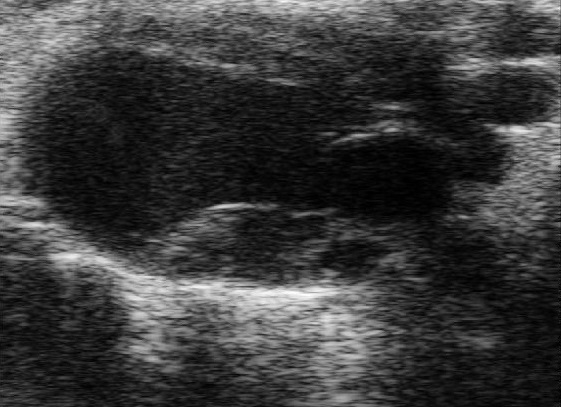}} \hspace{0.1cm}
\subfloat[ ]{\includegraphics[width=2.7cm, height=2.7cm]{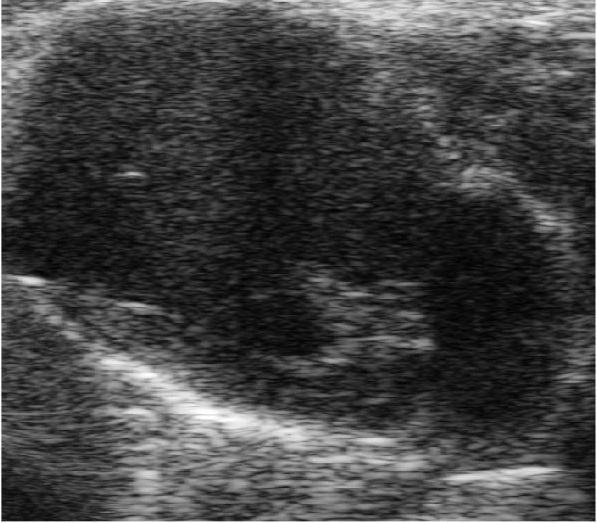}}\hspace{0.1cm}
\subfloat[ ]{\includegraphics[width=2.7cm, height=2.7cm]{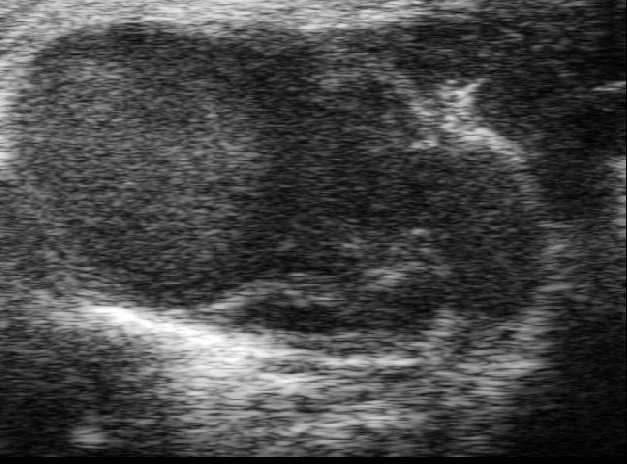}}\hspace{0.1cm}
\subfloat[ ]{\includegraphics[width=2.7cm, height=2.7cm]{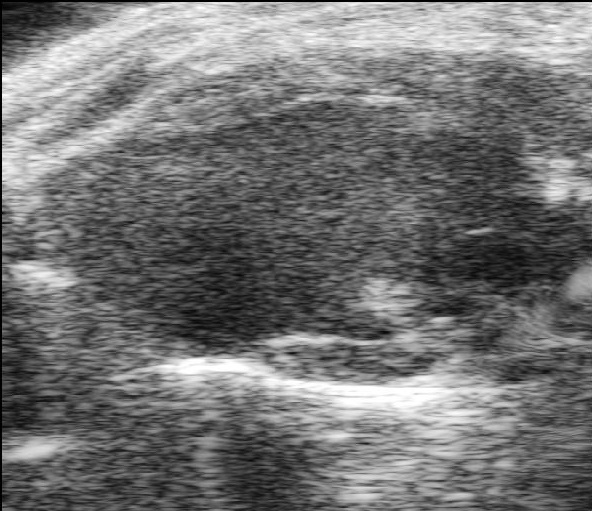}}\hspace{0.1cm}
\subfloat[ ]{\includegraphics[width=2.7cm, height=2.7cm]{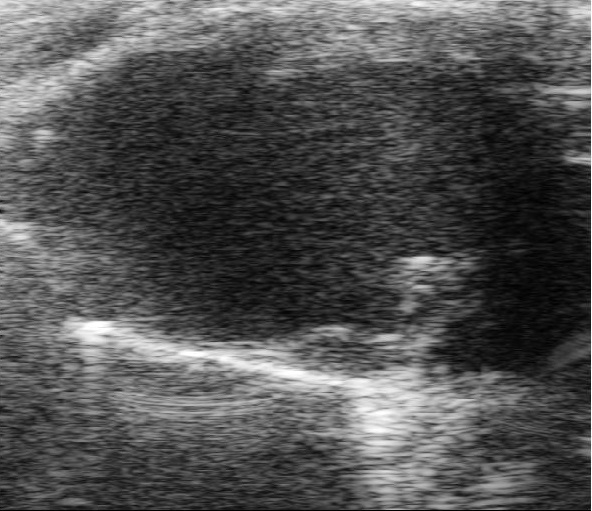}}\\

\rotatebox{90}{\hspace{1.1cm}\textbf{Ob}}\hspace{0.1cm}
\subfloat[ ]
{\includegraphics[width=2.7cm, height=2.7cm]{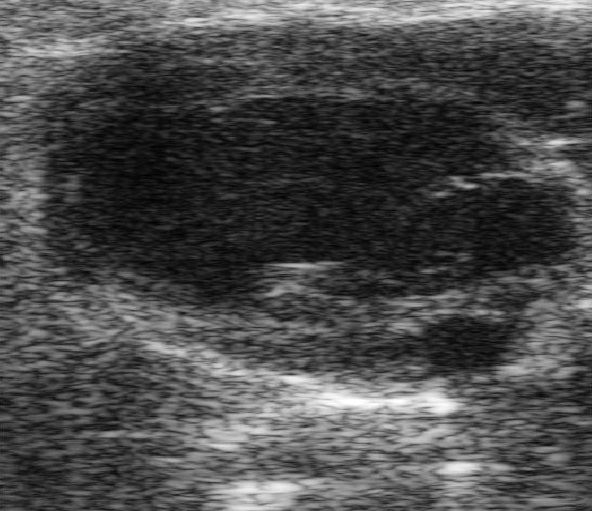}} \hspace{0.1cm}
\subfloat[ ]{\includegraphics[width=2.7cm, height=2.7cm]{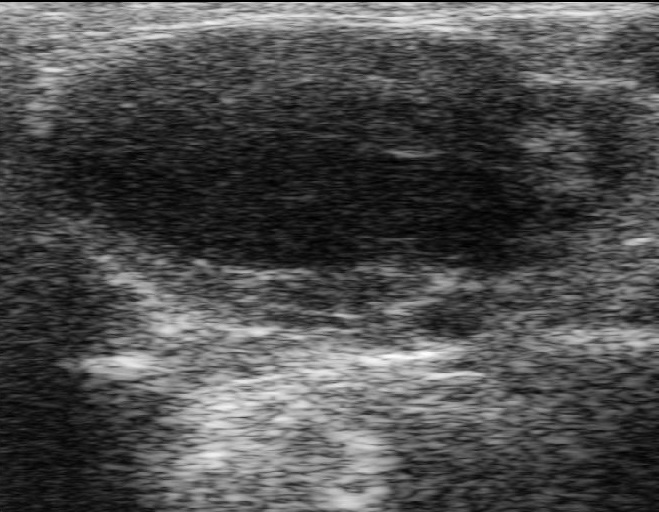}}\hspace{0.1cm}
\subfloat[ ]{\includegraphics[width=2.7cm, height=2.7cm]{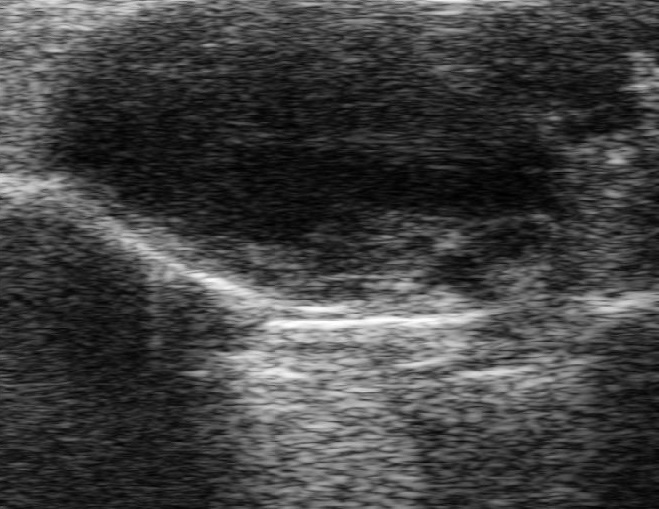}}\hspace{0.1cm}
\subfloat[ ]{\includegraphics[width=2.7cm, height=2.7cm]{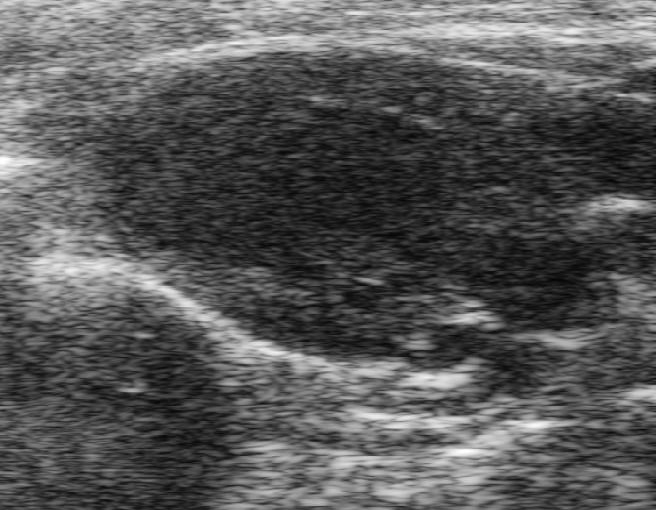}}\hspace{0.1cm}
\subfloat[ ]{\includegraphics[width=2.7cm, height=2.7cm]{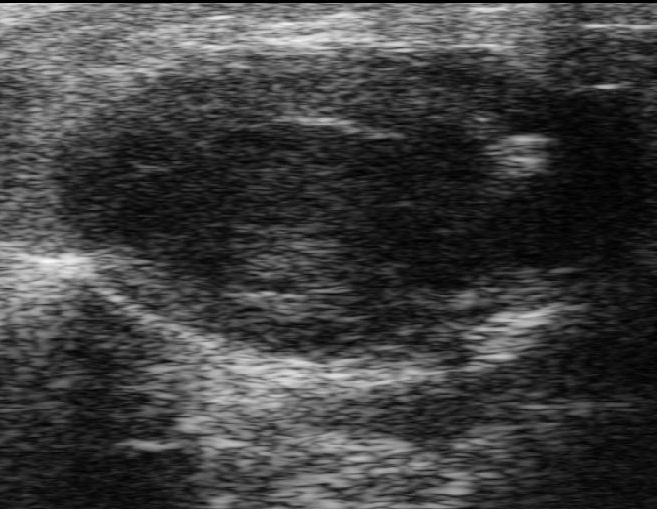}}\\

\rotatebox{90}{\hspace{1.1cm}\textbf{HT}}\hspace{0.1cm}
\subfloat[ ]
{\includegraphics[width=2.7cm, height=2.7cm]{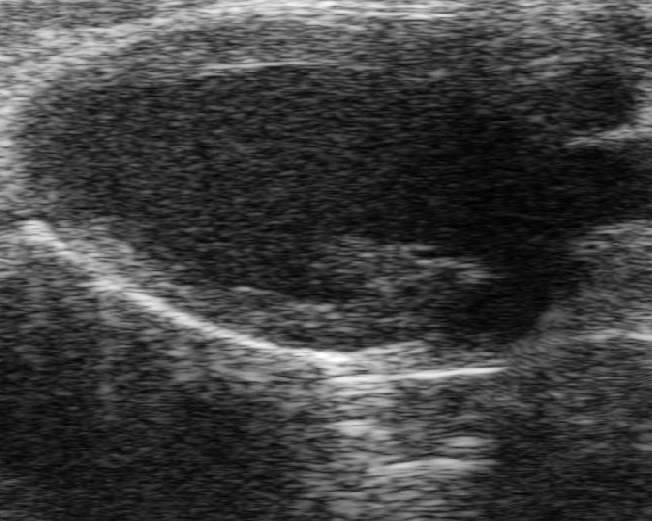}} \hspace{0.1cm}
\subfloat[ ]{\includegraphics[width=2.7cm, height=2.7cm]{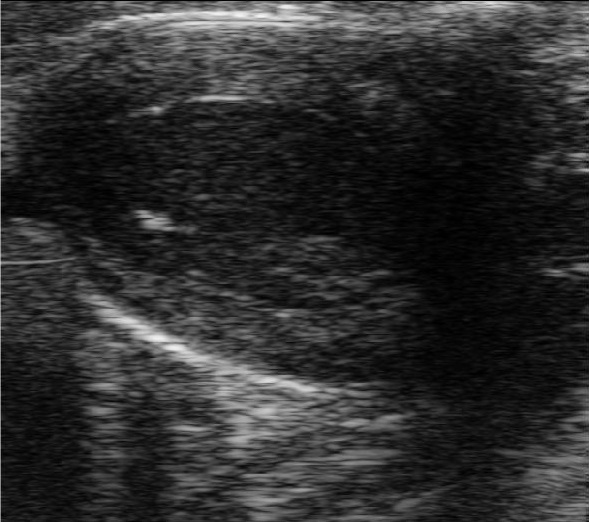}}\hspace{0.1cm}
\subfloat[ ]{\includegraphics[width=2.7cm, height=2.7cm]{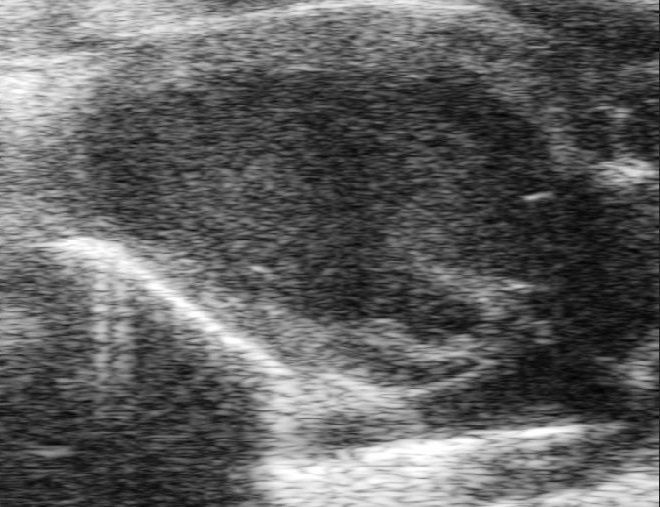}}\hspace{0.1cm}
\subfloat[ ]{\includegraphics[width=2.7cm, height=2.7cm]{figures/LAX23_DbM_007.jpg}}\hspace{0.1cm}
\subfloat[ ]{\includegraphics[width=2.7cm, height=2.7cm]{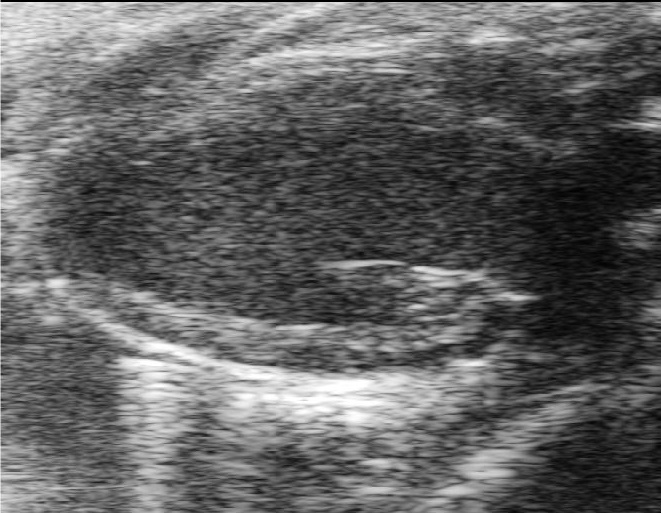}}\\

\centering
	\caption{Original echocardiography images. H: healthy, DC: diabetic cardiomyopathy, MI: myocardial infarction, Ob: obesity, HT: TAC hypertrophy }	
	\label{Fig06}
\end{figure}

\clearpage
\newpage

\begin{figure*}

	\centering
	\textbf{\hspace{0.3cm} Dataset 01 \hspace{1cm}  Dataset 02 \hspace{1.2cm}  Dataset 03  \hspace{1.2cm}  Dataset 04 \hspace{1.2cm}  Dataset 05 }\\

\rotatebox{90}{\hspace{0.8cm}\textbf{H}}\hspace{0.0001cm}
\subfloat[  ]{\includegraphics[width=2.7cm, height=2.7cm]{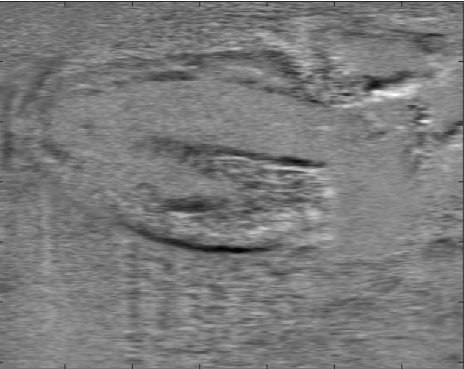}} \hspace{0.1cm} 
\subfloat[ ]{\includegraphics[width=2.7cm, height=2.7cm]{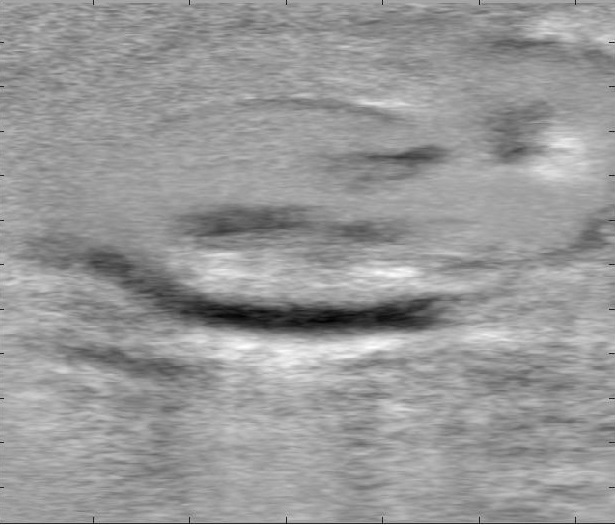}}\hspace{0.1cm} 
\subfloat[ ]{\includegraphics[width=2.7cm, height=2.7cm]{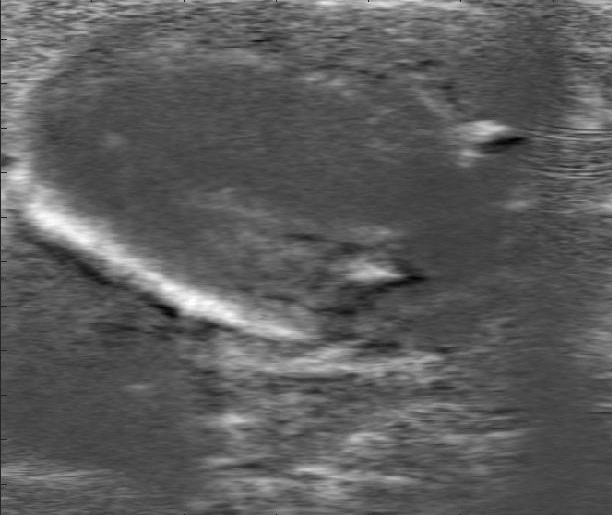}}\hspace{0.1cm} 
\subfloat[  ]{\includegraphics[width=2.7cm, height=2.7cm]{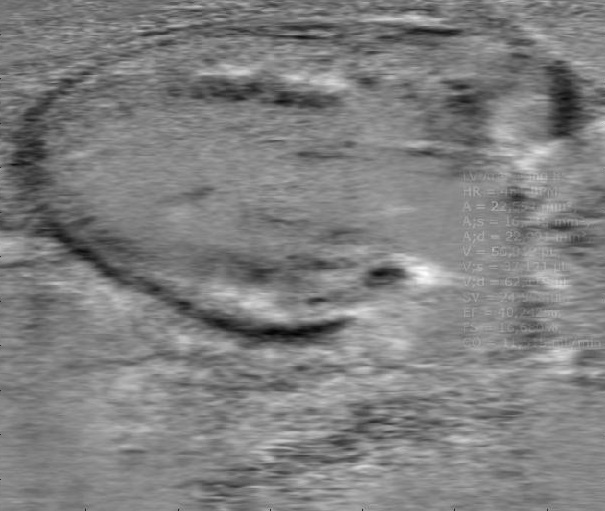}}\hspace{0.1cm}
\subfloat[ ]{\includegraphics[width=2.7cm, height=2.7cm]{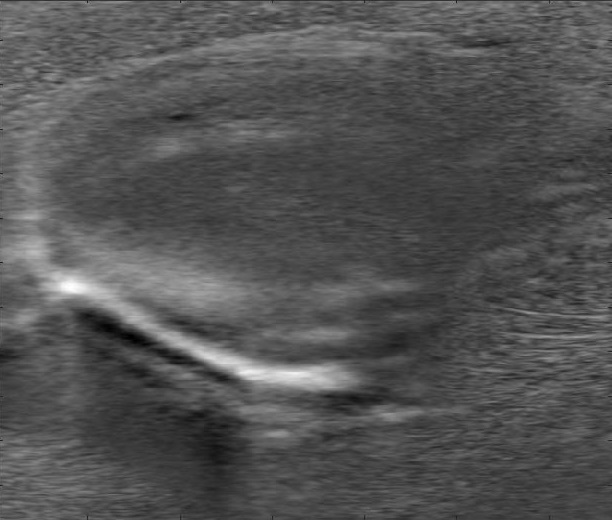}} 
\\

\rotatebox{90}{\hspace{1.1cm}\textbf{DC}}\hspace{0.1cm}
\subfloat[  ]{\includegraphics[width=2.7cm, height=2.7cm]{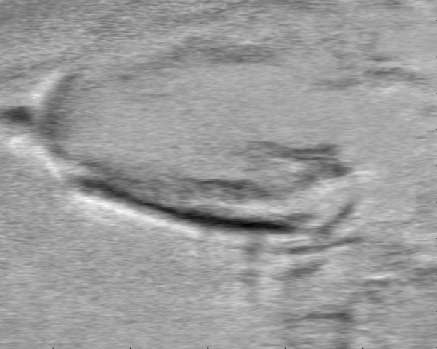}} \hspace{0.1cm} 
\subfloat[ ]{\includegraphics[width=2.7cm, height=2.7cm]{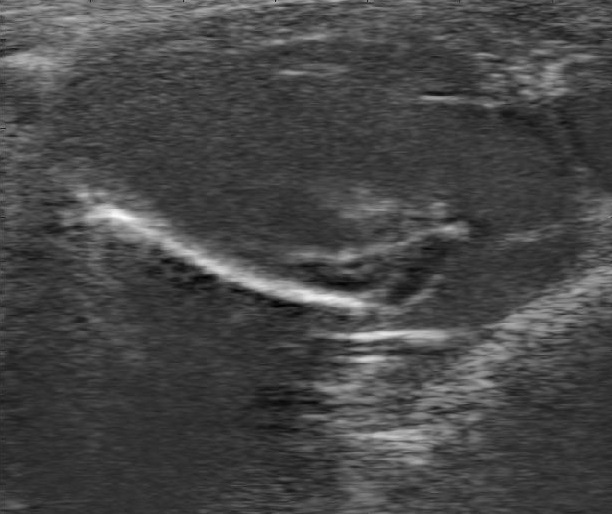}}\hspace{0.1cm}
\subfloat[ ]{\includegraphics[width=2.7cm, height=2.7cm]{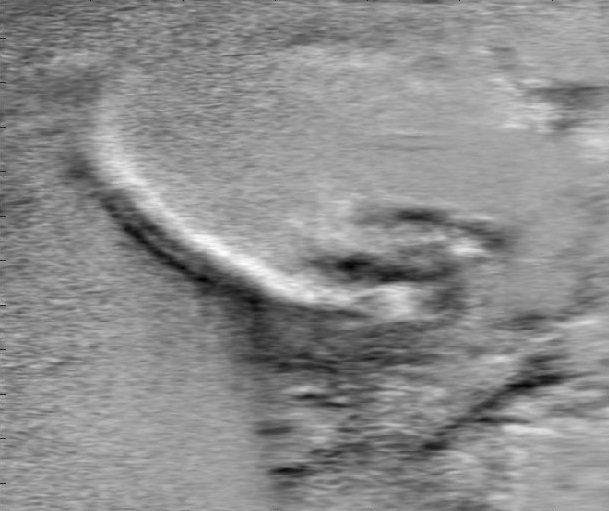}}\hspace{0.1cm}
\subfloat[ ]{\includegraphics[width=2.7cm, height=2.7cm]{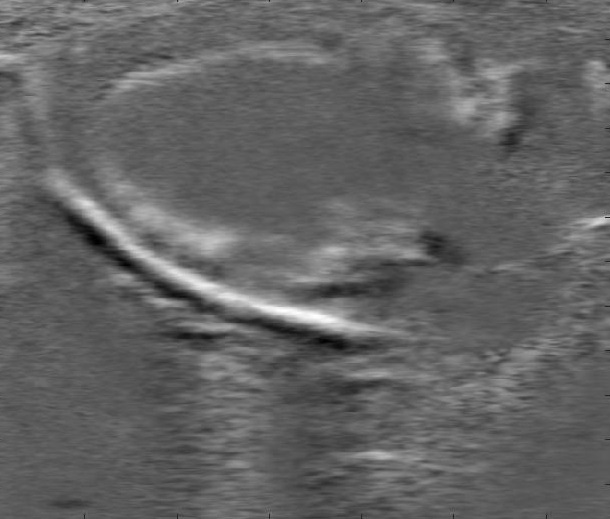}}\hspace{0.1cm}
\subfloat[ ]{\includegraphics[width=2.7cm, height=2.7cm]{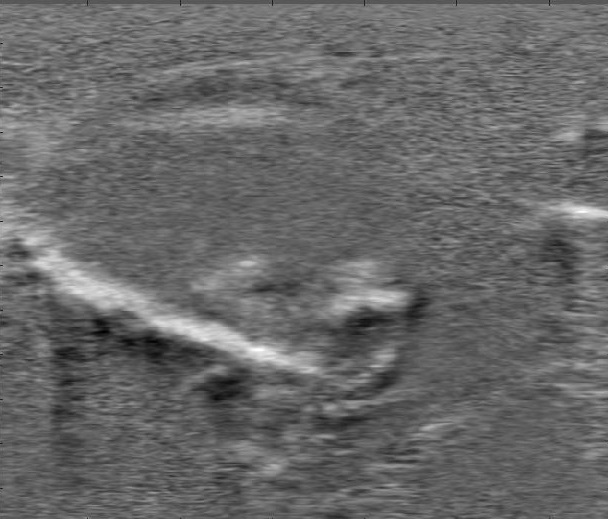}}\\

\rotatebox{90}{\hspace{1.1cm}\textbf{MI}}\hspace{0.1cm}
\subfloat[ ]
{\includegraphics[width=2.7cm, height=2.7cm]{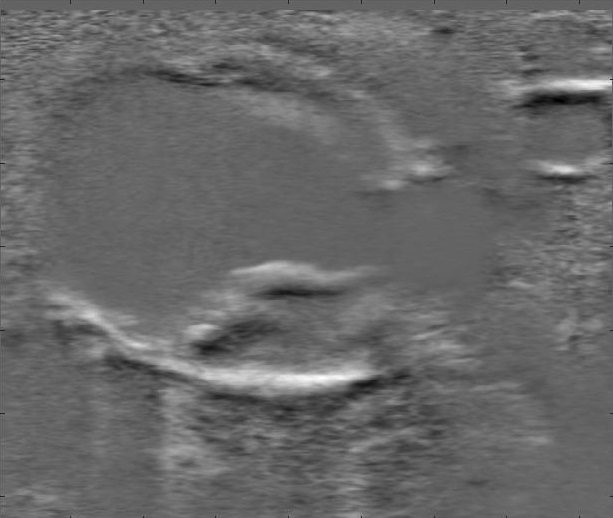}} \hspace{0.1cm}
\subfloat[ ]{\includegraphics[width=2.7cm, height=2.7cm]{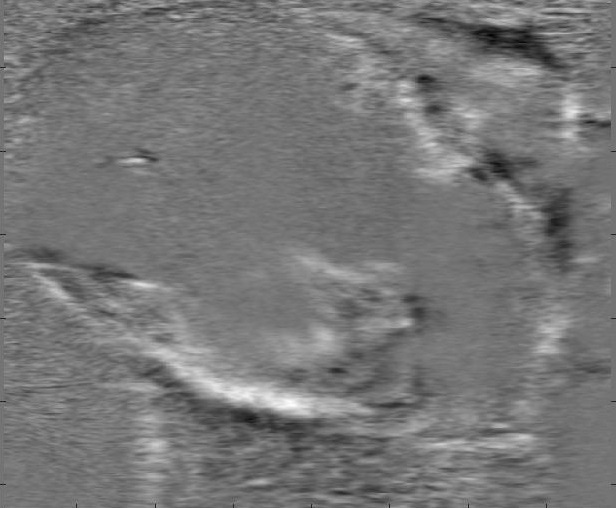}}\hspace{0.1cm}
\subfloat[ ]{\includegraphics[width=2.7cm, height=2.7cm]{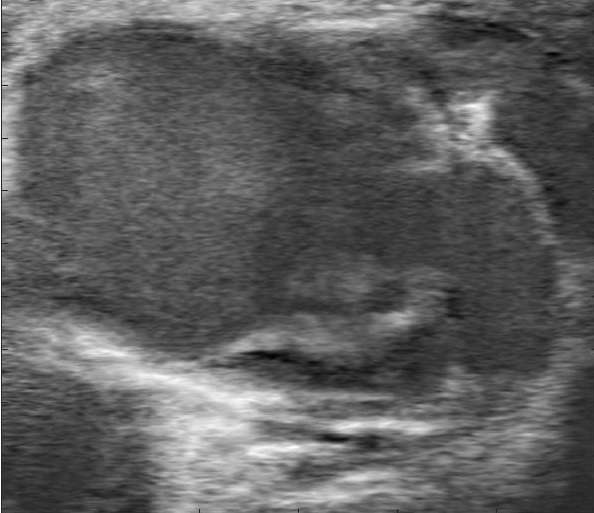}}\hspace{0.1cm}
\subfloat[ ]{\includegraphics[width=2.7cm, height=2.7cm]{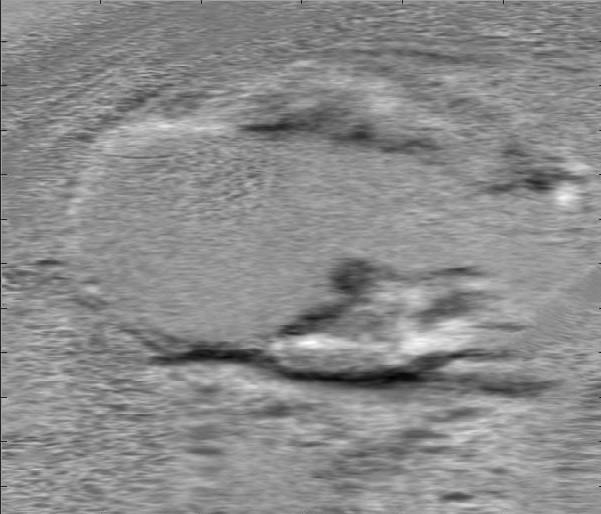}}\hspace{0.1cm}
\subfloat[ ]{\includegraphics[width=2.7cm, height=2.7cm]{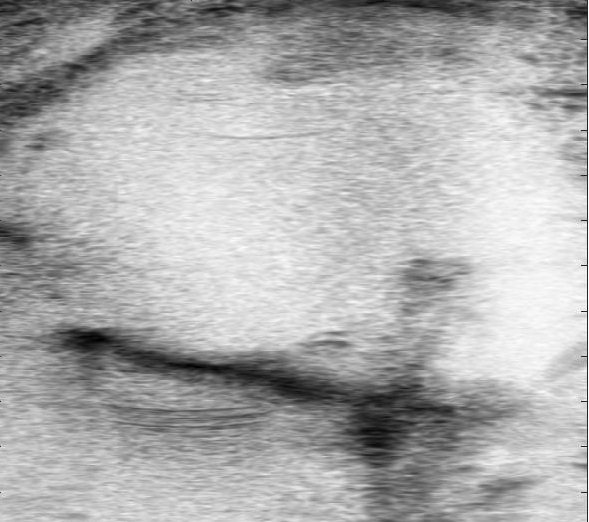}}\\

\rotatebox{90}{\hspace{1.1cm}\textbf{Ob}}\hspace{0.1cm}
\subfloat[ ]
{\includegraphics[width=2.7cm, height=2.7cm]{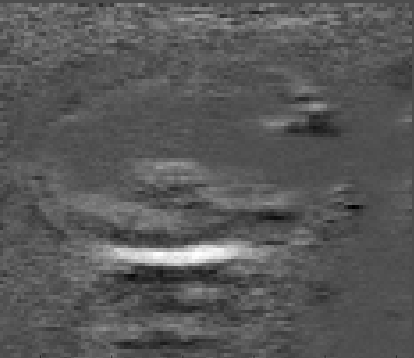}} \hspace{0.1cm}
\subfloat[ ]{\includegraphics[width=2.7cm, height=2.7cm]{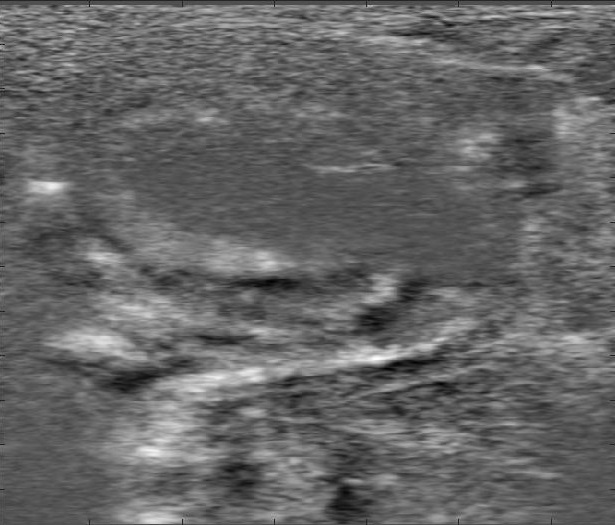}}\hspace{0.1cm}
\subfloat[ ]{\includegraphics[width=2.7cm, height=2.7cm]{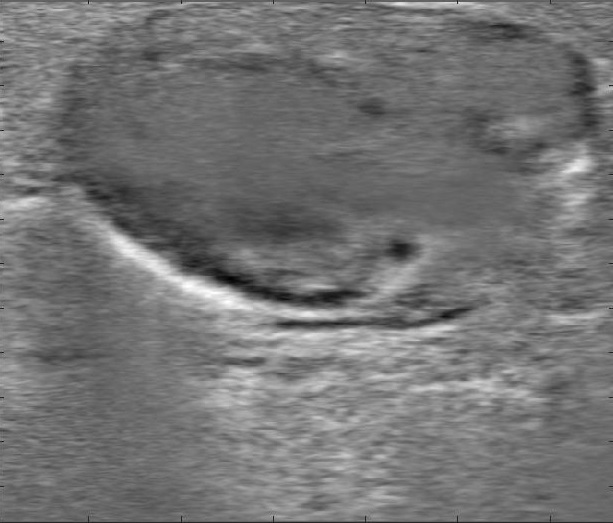}}\hspace{0.1cm}
\subfloat[ ]{\includegraphics[width=2.7cm, height=2.7cm]{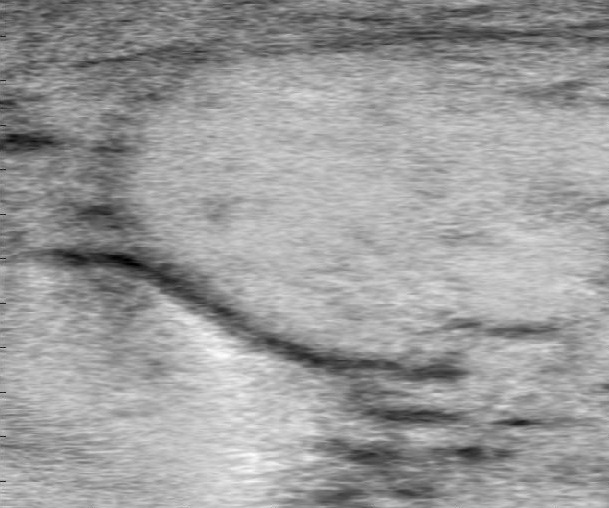}}\hspace{0.1cm}
\subfloat[ ]{\includegraphics[width=2.7cm, height=2.7cm]{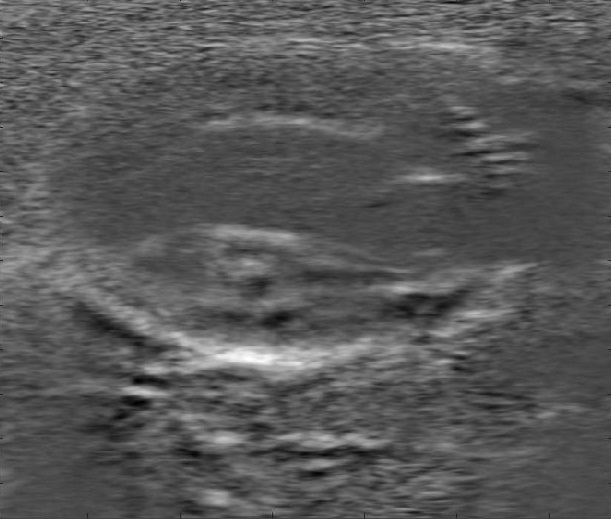}}\\

\rotatebox{90}{\hspace{1.1cm}\textbf{HT}}\hspace{0.1cm}
\subfloat[ ]
{\includegraphics[width=2.7cm, height=2.7cm]{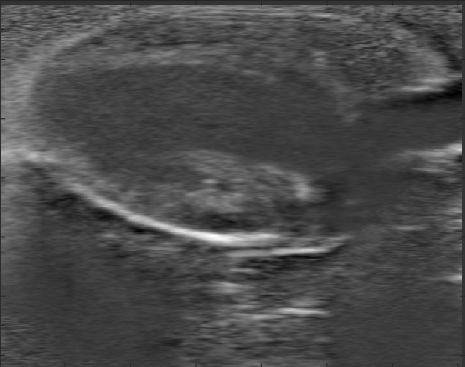}} \hspace{0.1cm}
\subfloat[ ]{\includegraphics[width=2.7cm, height=2.7cm]{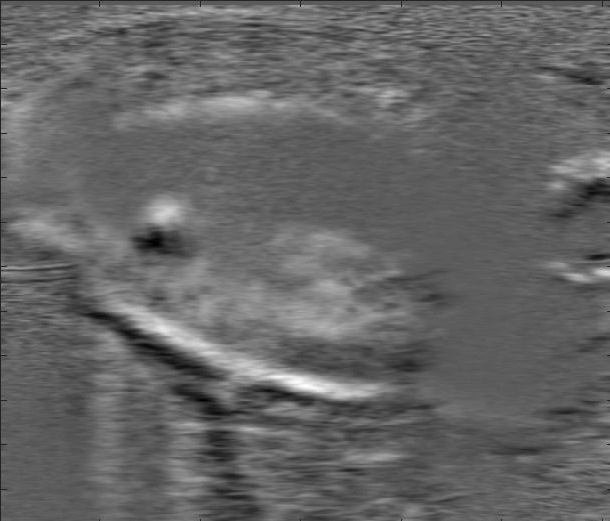}}\hspace{0.1cm}
\subfloat[ ]{\includegraphics[width=2.7cm, height=2.7cm]{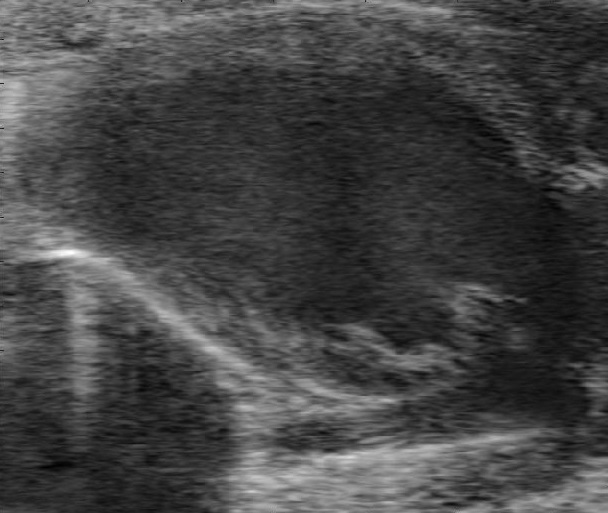}}\hspace{0.1cm}
\subfloat[ ]{\includegraphics[width=2.7cm, height=2.7cm]{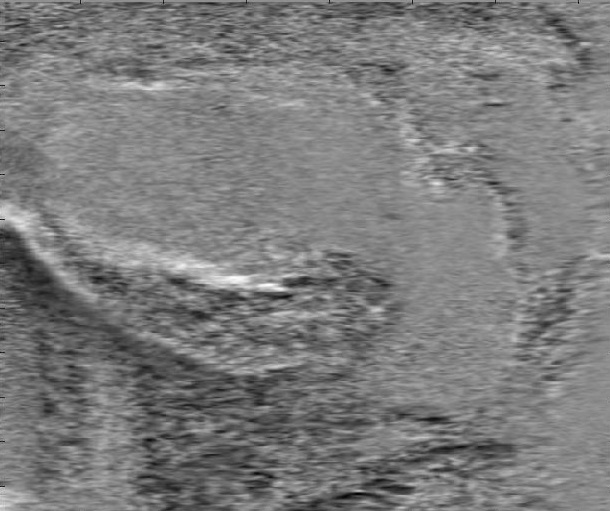}}\hspace{0.1cm}
\subfloat[ ]{\includegraphics[width=2.7cm, height=2.7cm]{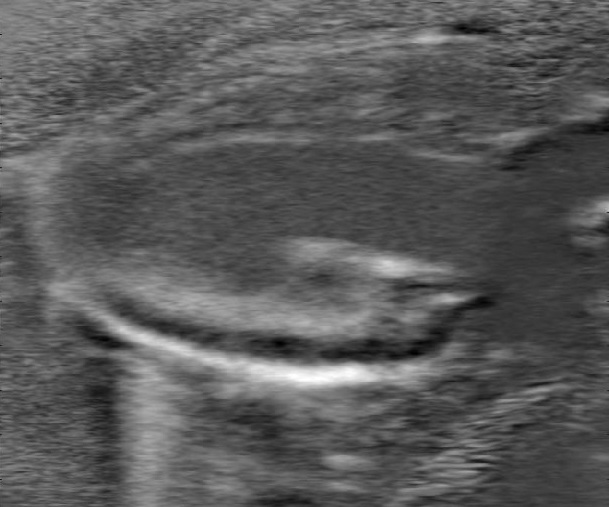}}\\

\centering
	\caption{Extracted features: DMD modes obtained using the HODMD algorithm. H: healthy, DC: diabetic cardiomyopathy, MI: myocardial infarction, Ob: obesity, HT: TAC hypertrophy }	
	\label{Fig07}
\end{figure*}
\FloatBarrier


\twocolumn
 \bibliographystyle{elsarticle-num} 
 \bibliography{My_Refs}





\end{document}